# Optimizing Drug Design by Merging Generative AI With Active Learning Frameworks

Isaac Filella-Merce[1,#], Alexis Molina[2,#], Marek Orzechowski[3,4], Lucía Díaz[2], Yang Ming Zhu[3,4], Julia Vilalta Mor[1], Laura Malo[2], Ajay S Yekkirala[3,4,$], Soumya Ray[3,6,$], Victor Guallar[1,5,$]

[1] Barcelona Supercomputing Center (BSC), Plaça d'Eusebi Güell, 1-3, 08034, Barcelona, Spain

[2] Nostrum Biodiscovery S.L., Av. de Josep Tarradellas, 8-10, 3-2, 08029, Barcelona, Spain

[3] RA Capital, 200 Berkeley Street, Boston Massachusetts 02116, United States

[4] Superluminal Medicines, 200 Berkeley Street, Boston Massachusetts 02116, United States

[5] ICREA, Pg. Lluis Companys 23, 08010, Barcelona, Spain

[6] 3-Dimensional Consulting, 134 Franklin Ave, Quincy MA-02170

#Contributed equally to this work.
$ corresponding authors (victor.guallar@bsc.es, ssray2000@icloud.com, ajay.yekkirala@superluminalrx.com)

# Abstract

Traditional drug discovery programs are being transformed by the advent of machine learning methods. Among these, Generative AI methods (GM) have gained attention due to their ability to design new molecules and enhance specific properties of existing ones. However, current GM methods have limitations, such as low affinity towards the target, unknown ADME/PK properties, or the lack of synthetic tractability. To improve the applicability domain of GM methods, we have developed a workflow based on a variational autoencoder coupled with active learning steps. The designed GM workflow iteratively learns from molecular metrics, including drug likeliness, synthesizability, similarity, and docking scores. In addition, we also included a hierarchical set of criteria based on advanced molecular modeling simulations during a final selection step. We tested our GM workflow on two model systems, CDK2 and KRAS. In both cases, our model generated chemically viable molecules with a high predicted affinity toward the targets. Particularly, the proportion of high-affinity molecules inferred by our GM workflow was significantly greater than that in the training data. Notably, we also uncovered novel scaffolds significantly dissimilar to those known for each target. These results highlight the potential of our GM workflow to explore novel chemical space for specific targets, thereby opening up new possibilities for drug discovery endeavors.

# Introduction

Machine learning (ML) based prediction methods are clearly beginning to contribute to, and in some cases even disrupt, the more traditional way of running drug discovery programs. Currently, they are being implemented in almost every phase of the early discovery process - from target validation, structure(s) generation, screening campaigns to ADME and toxicity predictions (1, 2). While the majority of ML methods in drug discovery still follow the 'design first and then predict' paradigm, more recently, generative AI methods (GM) are starting to attract significant attention (3, 4). GM methods are trained on large databases of chemical structures/properties to learn patterns and relationships between them. Once trained, these models can design new molecules presumably with specific properties, such as enhanced efficacy at a particular target (5) or minimizing side effects (6). Moreover, there is significant potential for new intellectual property as the designed novel compounds can be quite differentiated from existing inhibitors and those from a typical virtual screening campaign interrogating a large, but already enumerated, chemical space. Obvious disadvantages of introducing novel compounds via GM methods, might include difficulties in synthetic tractability, lack of target engagement, unknown ADME/PK properties and, importantly, the lack of efficiency of the designed compounds; as in other ML techniques, the applicability domain remains an open issue.

The most common types of GMs include Variational Autoencoders (VAEs), Generative Adversarial Networks and Autoregressive models. VAEs are a type of deep learning algorithm used for unsupervised learning, data compression and generative modelling, being composed of two main parts: an encoder network and a decoder network. The encoder network takes in the input data and compresses it into a lower-dimensional representation, the latent space, by mapping the input data to a probability distribution. The decoder network takes a sample from the latent space and reconstructs the original input data. The goal of the VAE is to learn the parameters that maximize the probability of generating the original input data. By doing so, VAEs learn to generate realistic samples in the latent space that can be used to infer novel data.

Even though there is no perfect framework for generative modelling, one might be inclined to use VAEs for several reasons. VAEs are known for their continuous and structured latent space, which allows for smooth interpolation between samples and better control over generated outputs (7). This is particularly useful for tasks requiring the generation of samples with specific properties or attributes, such as in drug design, where generating molecules with desired properties is crucial. Also, they are generally more stable and easier to train compared to Generative Adversarial Networks (GANs), as they have a well-defined optimization objective and do not suffer from issues like mode collapse or training instability (8). This ensures a more predictable and manageable training process. Moreover, VAEs have the ability to be forced to learn in low data regimes, whereas autoregressive models need to learn from lots of samples to produce meaningful results.

As stated, and similar to other ML techniques, there is often a lack of performance for molecules largely departing from the training set, the applicability domain problem. To address this issue, one might use an active learning procedure, where the GM model iteratively learns from the validation of its predictions (9, 10). Ideally, the input data sources would come from experimental validation, however, such implementation would drastically delay the overall design cycle. We find active learning in a virtual space with both physics-based and data-driven prediction techniques to address the various drug design parameters to be far more efficient. In an active learning centric workflow, by using synthesis accessibility predictors, drug-likeliness and results from physics based molecular modelling (MM) methods such as docking scores, we can continually evaluate the performance of the designed molecules and feed the information back to the GM. In general, virtual screening campaigns have shown that docking algorithms perform better with tight binders relative to weak binders thus compounds with improved docking scores tend to have less false positives. Thus by training the model on how to produce novel, better looking and high scoring molecules, we anticipate that the number of false positives will decrease significantly. Simply put, we aim to generate a list of drug-like molecules with (extremely) good docking scores, a high degree of novelty and large diversity. In addition, we propose not only to use MM techniques for the active learning but we also advocate to use them in the last ranking stages. Using a similar line of thought, if we have our list of virtual hits with high docking scores, we expect that more robust modelling techniques will further discriminate potential outliers. Therefore, we propose to run alternative modelling techniques, such as using Monte Carlo or molecular dynamics, to further narrow down the selection by applying consensus scores.

We have studied VAEs' active learning using two model systems with significantly different amounts of initial training data. The first target was the cyclin-dependent kinase 2 (CDK2), which drives the progression of cells into the S- and M-phases of the cell cycle. Although the role of CDK2 in tumorigenesis has been controversial, emerging evidence proposes that selective CDK2 inhibition may provide a therapeutic benefit against certain tumors (11, 12). First CDK2 inhibitors were reported as early as 2004 (13) and till date >10,000 known inhibitors have been disclosed against this target; several inhibitors have progressed to clinical trials (11, 14, 15), although an exquisitely CDK2-selective inhibitor is yet to be discovered. The second system involved was the kirsten rat sarcoma viral oncogene homologue (KRAS), which is one of the best-known oncogenes due to its high mutation rate in human cancer. KRAS has been largely associated with fatal cancers, including pancreatic (16), lung (17), and colorectal cancers (18). With the discovery of SII allosteric site (19), several irreversible covalently binding inhibitors of $KRAS^{G12C}$ have emerged, raising the hope of selectively drugging KRAS (20, 21). Most compounds currently under development, however, are tightly clustered around a single scaffold which was originally disclosed by Amgen (22). Recent studies have shown that SII can be employed to inhibit not only $KRAS^{G12C}$ but also $KRAS^{G12D}$ (23). This was achieved through the formation of a salt-bridge with D12 and an induced fit pocket to SII. Particularly, Mirati's research team discovered MRTX1133, a non-covalent inhibitor that can bind to the inactive and activated states of $KRAS^{G12D}$, resulting in KRAS pathway inhibition (24). Thus, from a generative AI perspective, we chose these two targets in an effort to evaluate the performance of our GM workflow against a densely populated patent space (CDK2) as well as a sparsely populated chemical space (KRAS).

# Results

## Designing a optimal generation workflow

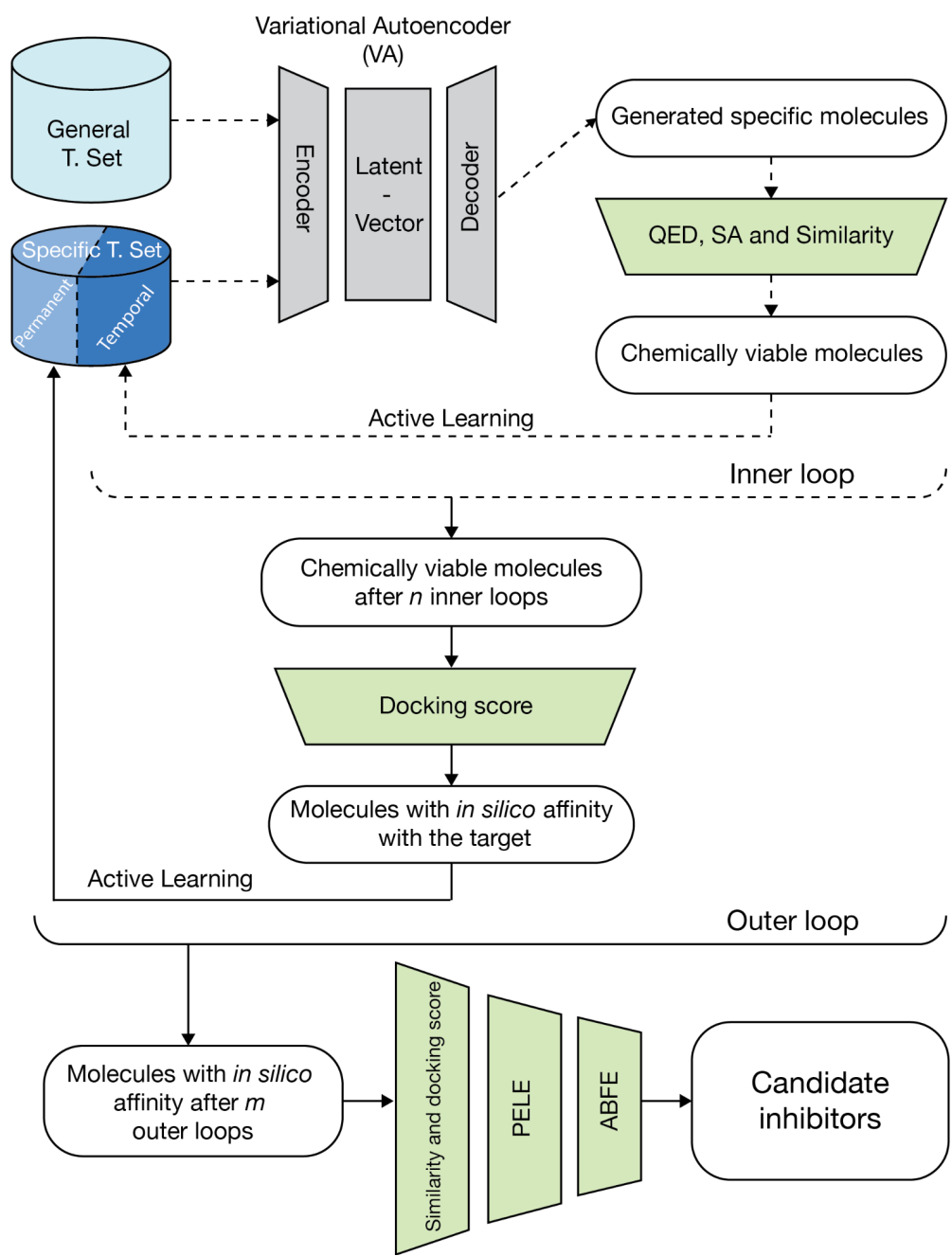

**Figure 1. Generation by active learning and selection workflow.** The GM workflow involves two nested iterative processes: a lower-level iteration (inner loop) and a higher-level iteration (outer loop). During each inner loop, new molecules are generated and filtered based on QED, SA score, and similarity. The resulting filtered molecules are used to enrich the temporal-specific set during the inner active learning step. Upon completion of *n* inner

loops, an outer loop filters the molecules from the temporal-specific set based on their docking score. The filtered molecules are then transferred from the temporal-specific set to the permanent-specific set via the outer active learning step. Thus, recall the two different levels of active learning in the overall workflow. After $m$ outer loops, all generated molecules in the permanent-specific set undergo further filtration using more stringent similarity and docking score thresholds. Finally, higher-level molecular modelling simulations are used to rank and select candidate inhibitors.

The GM workflow is guided by key insights and involves several steps. Initially, each molecular SMILES from the general training set (~250K from ChEMBL 30, see Methods) is input into a VAE to teach it how to generate viable chemical molecules. Next, the VAE's latent space is fine-tuned with molecular SMILES from an initial-specific training set, consisting of molecules with known or predicted affinity towards the target protein; for example, top scored compounds from a virtual screening might be used. This process is aimed at starting to teach the GM workflow to produce molecules with affinity towards a target protein. During this phase, a nested iterative process began with two active learning steps applied at two different stages within the GM workflow: inner loops and outer loops (Figure 1). During the inner loop process, the inner active learning step is crucial for enhancing the quality of generated molecules. In the outer loops, the outer learning step is aimed at directing the VAE to generate molecules with improved affinity towards the target protein.

In each inner loop, the GM workflow generates novel molecules, which are subsequently subjected to filtration by the inner active learning step. This evaluates each generated molecule based on QED, SA, and maximum similarity against the permanent-specific set. Thus, the purpose of the inner active learning step is to ensure that only molecules meeting drug-likeness, synthesizability, and variability requirements are added to a temporal-specific set, which is then used to fine-tune the VAE's latent space iteratively. Inner loops are crucial for improving molecular candidates during drug discovery, enabling effective exploration of chemical space and selection of potential drug candidates with desirable properties.

The outer loop functions as a higher-level optimization and selection process, evaluating molecular candidates based on their protein-ligand docking scores to identify potential drug candidates with strong binding affinities for the target protein. Following a predefined number of inner loops, the molecules in the temporal-specific set undergo docking simulations, with metrics such as Glide score used to assess the quality of protein-ligand interactions. Molecules meeting predetermined threshold criteria are transferred to the permanent-specific set, ensuring subsequent outer loops focus on refining the selection of high-quality molecules. The outer loop process is repeated for a set number of iterations, with each iteration narrowing down the molecular candidates based on docking scores.

Finally, more stringent filtration and selection processes are employed to identify the most promising candidates. Advanced molecular modelling simulations, such as PELE (25, 26) and molecular dynamics absolute binding free energy simulations (ABFE) (27), provide an in-depth evaluation of binding interactions and stability within protein-ligand complexes.

## CDK2

To evaluate and optimize our GM workflow, we began by testing it with CDK2, which serves as an optimal starting point. This is due to the significant amount of data available on known CDK2 inhibitors and the need for novel inhibitors with unexplored CDK2 scaffolds that can surpass the promiscuity commonly linked with kinase inhibitors. In this regard, our GM workflow allowed us to control the scope of molecular exploration and variability through the similarity parameter included in the inner active learning step, as well as regulate the affinity towards CDK2 with the docking scores of the outer active learning step.

Due to the exploratory nature of this first test, we began by executing a single outer loop comprising 16 inner loops. For training, we used a general set of over 200k molecules and a CDK2 specific set of 1,061 distinct molecules with experimentally tested affinity to CDK2 (see Methods). In the inner active learning step, the QED, SA, and similarity thresholds were set at 0.6, 7, and 0.6, respectively. Upon completing the first outer loop, we generated 49,796 new molecules. Out of these, 15,068 satisfied the inner active learning thresholds, contributing to the enrichment of the temporal-specific set. Notably, the generation of these molecules was unevenly distributed throughout the process (Figure 1A); notice the sharp drop in generation for some iterations as a result of the stochastic nature. However, the average percentage of generated molecules passing the thresholds was 32.5% ± 14.4%. If we exclude the first and second inner loops, which had a percentage of accepted molecules of 72.3% and 48.6%, respectively, the average percentage drops to 28.5% ± 9.4%. This behavior was more consistent, as demonstrated by the semi-linear trend of the temporal-specific set enrichment plot over the inner loops (Figure 2B-C).

The two-dimensional molecular representation provided by the Uniform Manifold Approximation and Projection (UMAP) algorithm (28) allowed us to visualize the exploratory nature of our GM workflow over the inner loops (Figure 2D). During the first inner loop enrichment (inner1), the GM workflow discovered a couple of nearby regions while exhaustively exploring the neighboring of the initial-specific set. From the second (inner2) to the sixth (inner6) inner loop enrichments, the workflow continued to scout the two regions uncovered by inner1 and began exploring distant additional regions. Finally, from seventh (inner7) to sixteenth (inner16) inner loop enrichments, the GM workflow intensively explored the distant regions discovered by inner1 to inner6, generating molecules far from any molecule from the initial-specific set. This led to a burst in the exploration of diverse regions.

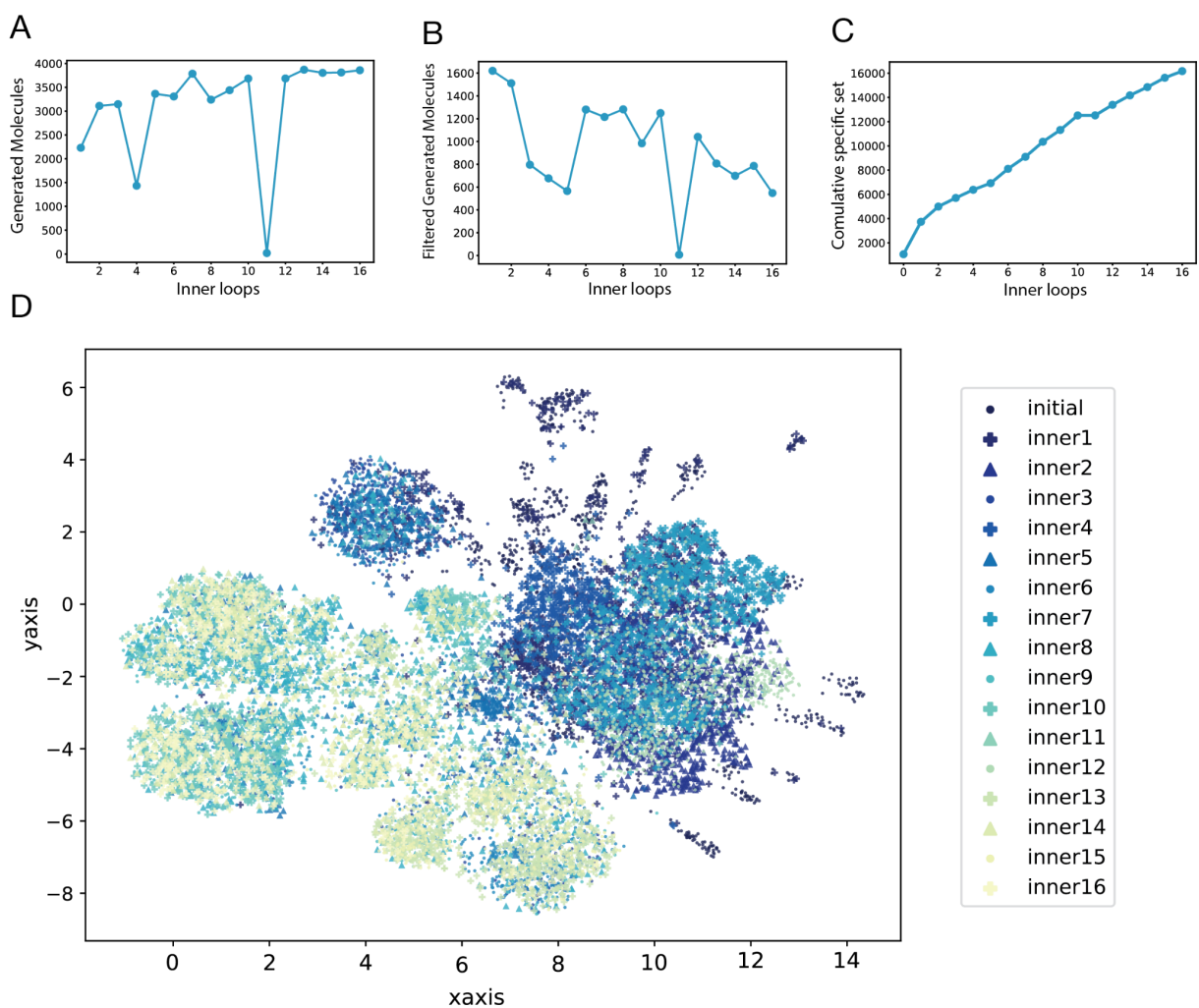

**Figure 2. Inner loop decomposition of the first outer loop for CDK2. A.** Total number of generated molecules at each inner loop. **B.** Number of generated molecules at each inner loop that meet the inner loop filters. **C.** Cumulative plot representing the enrichment of the temporal-specific set over the inner loops. **D.** UMAP illustrating the enrichment of the temporal-specific set over the inner loops. Marker's styles and colors of molecules from different inner loops are indicated in the legend. The initial-specific set is represented as 'initial' and the corresponding enrichments at each inner loop as 'innerX'.

The first outer loop was concluded by the outer active learning step, which involved calculating a docking score for each newly generated molecule in the temporal-specific set and filtering them based on a predefined threshold. Glide docking was utilized for this purpose, and a Glide gscore threshold of -8.0 was set to filter the molecules from the temporal-specific set to the permanent-specific set (see Methods). Out of the 15,068 molecules in the temporal-specific set, only 885 met the threshold and were thus transferred to the permanent-specific set for the next outer loop.

In the second outer loop, the GM workflow generated 24,766 molecules. Of these, 12,040 passed the inner active learning steps filters, and 1,192 passed the outer active learning step filter (Figure 3A). In the first outer loop, only 30.3% and 1.8% of the molecules passed the inner and outer active learning filters, respectively. However, in the second outer loop, the numbers increased to 48.6% and 4.8%, indicating that the GM workflow was learning to generate more diverse molecules with a greater affinity towards CDK2 (Suppl. Table 1). Despite the success of the first two outer loops, the third outer loop proved challenging for the GM workflow. It generated 26,387 molecules, but only 2,065 (7.8%) fulfilled the inner active learning step filters, and only 167 (0.6%) passed the outer learning step filters. This indicated that the GM workflow struggled to increase the variability of the generated molecules in the third outer loop. To address this variability problem, we ran a fourth independent outer loop, in which we again trained the model from the initial-specific set. However, this time, we used a more restrictive inner active learning similarity filter of 0.4. The GM workflow generated 19,581 molecules during this loop, of which only 1,446 (7.3%) passed the inner active learning filters (as expected due to the threshold decrease), and only 88 (0.5%) met the outer active learning filters. For the fifth and last outer loop, we added all the molecules that had passed the outer active learning filters of the third and fourth loops to the permanent-specific set. By doing so, the GM workflow generated 36,465 molecules, of which 18,645 (51.1%) passed the inner active learning filters, and 2,295 (6.3%) met the outer active learning filters. This approach helped us generate more diverse molecules with a higher affinity towards CDK2. The affinity improvement can be observed in Figure 3C-D. There we appreciate how within the Glide gscore range of -8.0 to -11.5, we generated 22 times more molecules than were present in the initial-specific set, increasing the number from 219 to 4,821 (from which 3,142 had a maximum similarity against the initial-specific set below 0.3). Notably, we generated 28 molecules with a Glide gscore below -11.5, a 9-fold increase compared to the 3 molecules present in the initial specific set. Out of the 28, 11 had a maximum similarity below 0.3 against the initial-specific set.

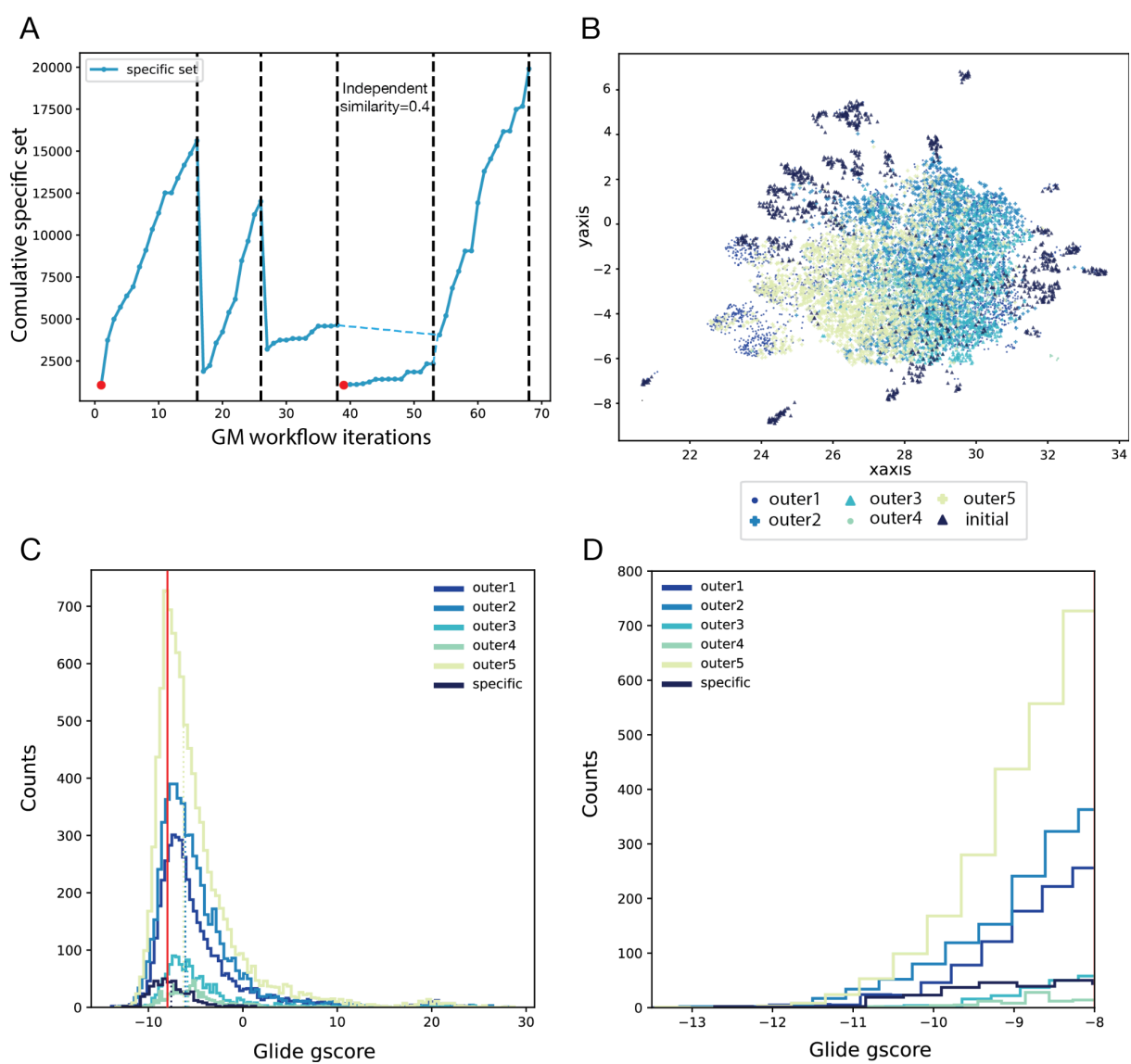

**Figure 3. Generated molecules for CDK2 over five outer loops. A.** Cumulative plot representing the enrichment of the CDK2 specific set over the inner and outer loops. The vertical dotted lines highlight each outer loop's last inner loop, indicating the plot's splitting into outer loops. Red dots indicate inner loops, whose specific set was the initial-specific set of CDK2. The blue dotted lines connecting non-consecutive outer loops indicate an outer active learning step between them. **B.** UMAP plot illustrating the generated molecules over the outer loops. The markers' styles and colors of the molecules from different outer loops are indicated in the legend. **C.** Histograms of the Glide gscore obtained during the outer active learning steps and the Glide gscores of the molecules from the initial-specific set of CDK2. The red vertical line represents the thresholds of -8.0 of gscore. **D.** Zoomed-in view of the best scoring compounds in panel C.

The UMAP of the outer loops (Figure 3B) provides a visual summary of the CDK2 generative process. The figure illustrates how molecules from the first outer loop (outer1) populated the neighboring space of the initial-specific set molecules (initial) and sparsely proliferated across the space. Hereafter, molecules from the second outer loop (outer2) exhaustively explored most of the regions sparsely explored by outer1, leaving the neighboring area of the initial-specific set. Next, molecules from the third outer loop (outer3) remained close to the outer1 and outer2 molecules, only exploring already densely populated regions. This reduced their probability of passing the similarity filters during the third outer active learning step. Notably, molecules from the fourth and independent outer loop (outer4) exhibited a different behavior than molecules from outer1, even though they shared the same initial-specific set. In contrast to outer1 molecules, outer4 ligands did not populate the neighboring space of the initial-specific set due to the more restrictive similarity filter. Instead, they sparsely explored the central region. Finally, molecules from the fifth outer loop (outer5) began exploring low, densely populated regions, thanks to the increase in variability caused by outer4. Overall, the UMAP in Figure 3B effectively summarizes the generative process of the outer loops of CDK2, highlighting the unique characteristics of each outer loop. Supplementary Figure 1A complements this visualization by using an alternative color scheme to clearly show the extent to which the GM workflow explored the chemical space beyond the CDK2 initial-specific set. Finally, Supplementary Figure 1B shows that the generated molecules that met the outer filters are evenly distributed throughout the newly explored chemical space.

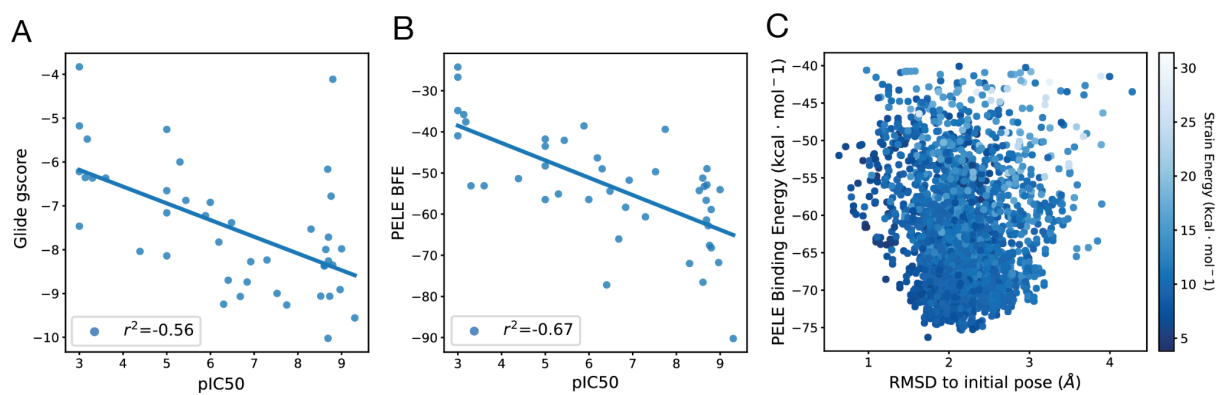

**Figure 4. Comparison of Glide and PELE simulations in ranking molecules from the initial-specific set of CDK2. A.** Correlation between Glide gscore and pIC50 computed from a subset of molecules from the initial-specific set of CDK2. **B.** Correlation between PELE's BFE and pIC50 computed from the same subset of molecules from panel A. **C.** Illustration of one of the PELE's energetic profiles employed to compute PELE's BFE values from panel B.

After all outer loops passed, the generated molecules that met the outer active learning filters underwent a final filter procedure to discard/rank them (Figure 1). These involved a more stringent similarity check (pose similarity) along with more robust MM simulations. As stated, we expect that robust MM simulations would further discriminate docking outliers, and that a consensus scheme might help in deciding the final list of candidates. To test this idea, we first sought to assess the potential of utilizing the all-atom Monte Carlo PELE software (29) to enhance the precision of docking scores. To do so, we ran PELE on a subset of CDK2 inhibitors from the initial-specific set. Specifically, we ran rescoring PELE simulations using the docking poses from Glide as a starting point. By comparing the Glide gscores and the PELE binding free energy estimations (BFE) against the experimental affinities, we determined the degree to which PELE improved the accuracy of the docking scores (Figure 4A-C). We saw a significant improvement in the correlation between the experimental and predicted values, with the correlation coefficient increasing from 0.56 to 0.67 from Glide to PELE. These findings suggest that (as expected) PELE could be a valuable tool for enhancing the accuracy of the Glide docking poses and scores of newly generated CDK2 molecules.

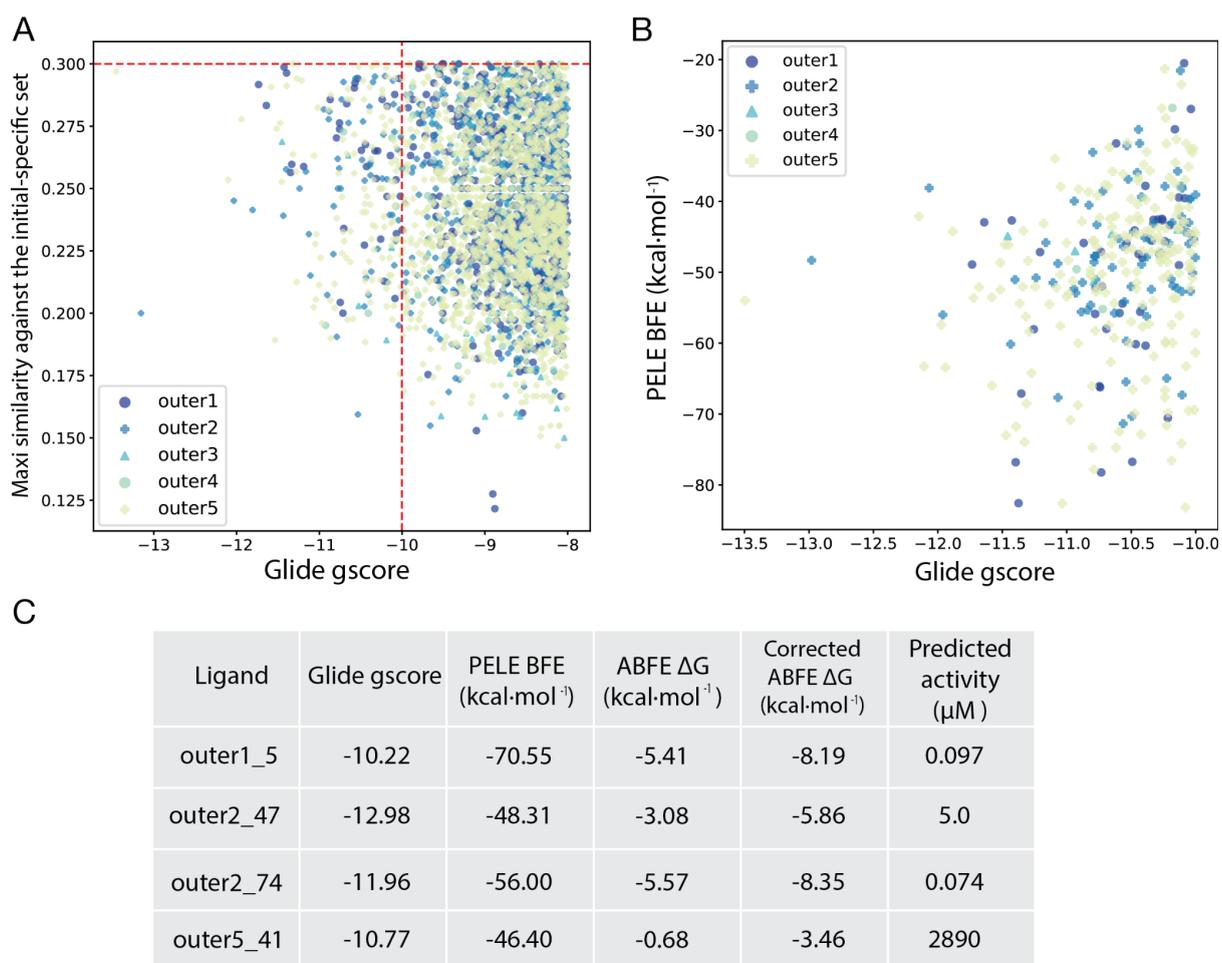

**Figure 5. Selection of CDK2 candidate inhibitors from the pool of newly generated molecules. A.** Scatter plot of all CDK2 generated molecules with a Glide gscore below -8 and a maximum similarity against the initial-specific set of 0.3. The red dotted vertical and horizontal lines represent the selection thresholds of -10 for gscore and 0.3 for maximum similarity against the initial-specific set, respectively. **B.** Scatter plot of PELE BFE and Glide gscore of molecules fulfilling panel A's gscore and similarity selection thresholds. **C.** Affinity predictions for the selected compounds fulfilling the selection thresholds from panel A and with an atypical hinge binder scaffold for CDK2.

PELE rescoring simulations (Figure 5B) were carried out on the generated molecules, fulfilling more stringent filters of Glide gscore and pose similarity than those applied in the active learning step. Specifically, we selected the 228 molecules with a Glide gscore and pose similarity below -10 and 0.3, respectively (Figure 5A). In addition, we confirmed that the selected molecules were evenly distributed throughout the newly explored chemical space by our GM workflow (Suppl. Figure 1C). Known ligands for the protein structure

3BHV were also used for benchmarking the performance of ABFE calculations with the CDK2 system. We observed that ABFE consistently underpredicted the binding affinity relative to the experimental ΔG by 2.08Kcal/mol. While we don't fully understand the origin of this error, we used 2.08Kcal/mol as a correction/offset applied to the other ligands measured using ABFE (Figure 5C). For this selection, and on top of the 0.3 pose similarity and -10 Glide gscore cutoffs, we applied an additional PELE and similarity filter. For PELE, we only looked at those molecules in the bottom half (with BFE lower than -40 kcal/mol). Then we visually inspected all bound molecules, after the induced fit from PELE, and selected those with a more dissimilar kinase hinge region, thus aiming to identify atypical CDK2 scaffolds. Figure 5C shows the affinity predictions for these four selected compounds, where we observe how the two more computationally intensive methods, PELE and ABFE, agree in prioritising the molecules, despite of using different modelling approaches (Monte Carlo and ABFE); the more approximate Glide docking procedure seems to fail in adequately ranking them.

## KRAS

We tested our GM workflow to generate molecules that could inhibit $KRAS^{G12D}$ through its SII allosteric site. This task was considerably more challenging than the CDK2 case due to the scarcity of known $KRAS^{G12D}$ inhibitors. We could only gather a set of 73 molecules with experimental tested affinity to KRAS, referred to as the KRAS known specific set, with 83% of these molecules targeting SII for $KRAS^{G12D}$ inhibition and the rest targeting SII for $KRAS^{G12C}$ inhibition (covalent inhibitors that can be used after removing their covalent warhead). These molecules were primarily developed using structure-based activity improvement strategies on a small number of scaffolds (24, 30), resulting in little variability within the KRAS known specific set. To anticipate potential issues due to the small number of molecules and their lack of variability, we also created an extra specific set of molecules with unknown experimental affinity, referred to as the KRAS unknown specific set. This additional specific set of 1,891 molecules was obtained through a high-throughput virtual screening (HTVS) on a subset of Enamine molecules (see Methods).

Using both specific sets, we conducted two generative processes in parallel, one starting from the known specific set and another from the unknown specific set. Each consisted of four outer loops, and each outer loop included 15 inner loops. We used QED, SA, and

Similarity thresholds of 0.6, 7, and 0.6, respectively, for the inner active learning steps and a glide gscore threshold of -8 for the outer active learning steps.

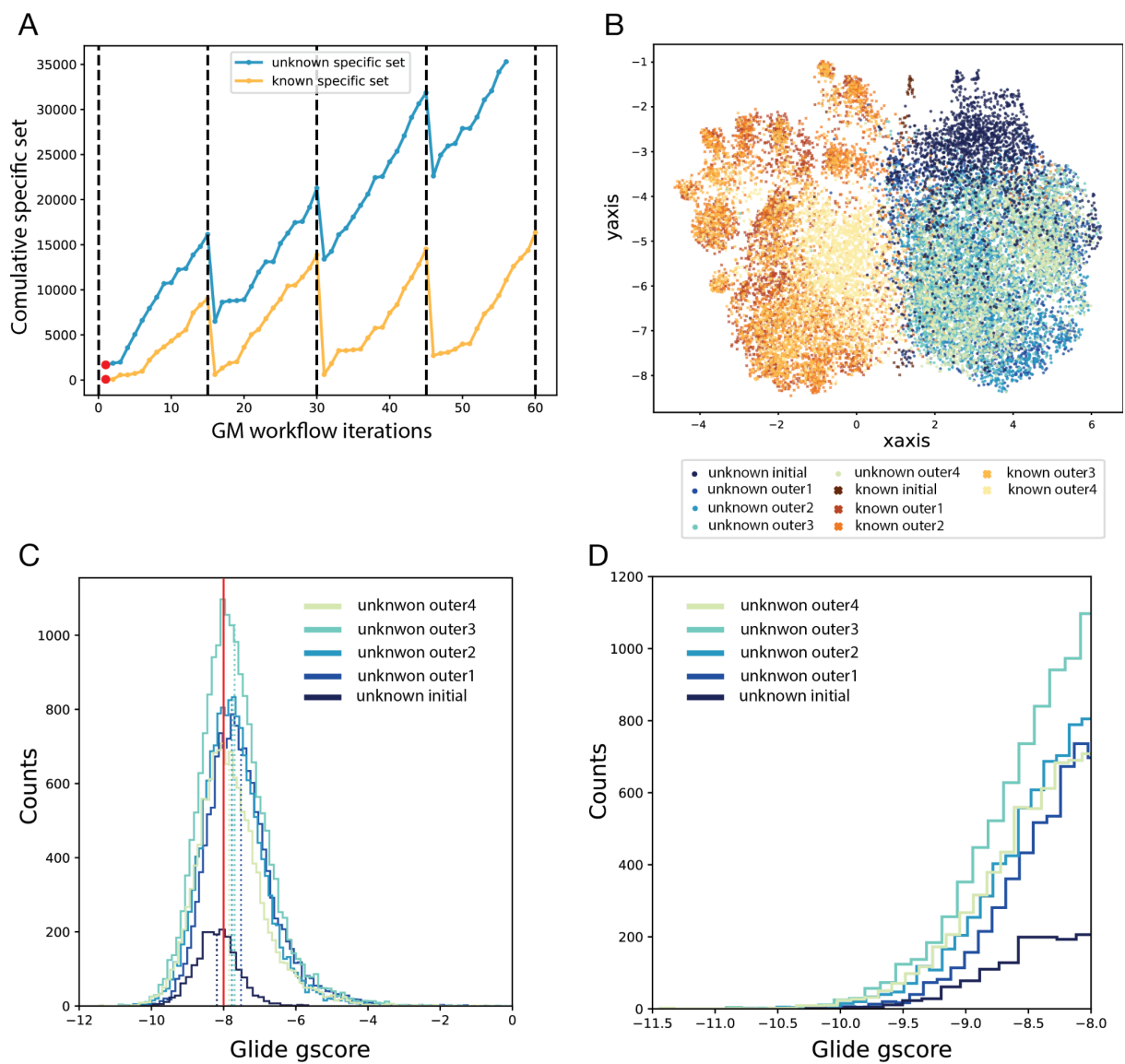

**Figure 6. Generated molecules for KRAS$^{G12D}$ with the known and unknown initial-specific sets over four outer loops. A.** Cumulative plots depicting the enrichment of the KRAS$^{G12D}$ known and unknown specific sets over the inner and outer loops. The vertical dotted lines highlight each outer loop's last inner loop, indicating the plot's splitting into outer loops. The red dots indicate inner loops whose specific set was the known or unknown initial-specific set of KRAS$^{G12D}$. **B.** UMAP plot illustrating the known and unknown generated molecules over their corresponding outer loops. Marker's styles and colors of molecules from different outer loops and different initial-specific sets are indicated in the legend. **C.** Histograms of the Glide gscores obtained during the outer active learning steps

for the generation process initiated with the unknown initial-specific set and the Glide gscores of the molecules from the unknown initial-specific set of KRAS$^{G12D}$. **D.** Zoomed-in view of the right tail of the histograms in panel C.

Compared to CDK2, the two KRAS generative processes were more consistent across the outer loops (Figure 3A and Figure 6A). These did not present an event such as the third outer loop of CDK2, where only 7.8% of the generated molecules met the inner active learning step filters. Therefore, there was no need to apply an independent outer loop, as we did on the fourth outer loop of CDK2. Notably, the average percentage of generated molecules fulfilling the inner and outer active learning steps during the KRAS known generation was similar to that of CDK2: 2.3±1.3% and 2.8±2.3%, respectively (Suppl. Table 2). On the contrary, this average increased to 15.8±2.9% on the KRAS unknown generation. The key difference between the CDK2 specific set and the KRAS known specific set, as compared to the KRAS unknown specific set, is the use of molecules with experimentally tested affinity against the usage of molecules obtained through an HTVS. The increase in molecular variability in the KRAS unknown specific set, as compared to the known specific set (as seen in Suppl. Figure 2), compensates for the loss of knowledge of experimental affinity and allows the GM workflow to obtain higher yields of molecules fulfilling the active learning steps.

The UMAP plot in Figure 6B clearly distinguishes between the molecules generated during the known and unknown generative processes of KRAS. Specifically, two distinct clusters represent the two processes. The known cluster contains almost no unknown molecules, while the unknown cluster contains only a few known molecules scattered throughout. However, the dispersed known molecules did not initiate a thorough exploration of the unknown cluster. In the known cluster, the molecules from the initial-specific set (known initial) are mainly located at the border between the two clusters. Instead, the molecules from the first outer loop (known 1) are mostly situated in groups on the left side of the UMAP. The molecules from the two subsequent outer loops (known 2 and known 3) extensively explored the neighboring of molecules from known 1 up to the emergence of a new group of molecules in the center of the UMAP, formed by the final outer loop molecules (known 4). On the other hand, the unknown cluster started its formation with molecules from the first outer loop (unknown 1), which sparsely explored most of the cluster's surface, starting from the top, where the initial-specific set molecules (unknown initial) are located. In the following outer loops (from unknown 1 to unknown 4), the unknown generative process thoroughly explored the unknown cluster's surface. To sum up, this UMAP depicts two distinct generative processes that yielded two unrelated sets of molecules.

To understand the differences that prompted the formation of the two unrelated clusters, we visually inspected their molecules. We found that most of the known molecules had anomalous chemical structures, such as broken bicycles or 7-membered rings or bigger (Suppl. Figure 3). These artifacts seem to arise from two positively charged bicycles decorating most of the KRAS scaffolds (such as the ones from MRTX1133 and derivates). We inferred that our GM workflow could not interpret such structures correctly, resulting in the generation of non-natural chemical structures. Given that the unknown molecules did not exhibit this issue, we proceeded with the analysis using only the unknown molecules. Importantly, we observed a significant improvement in affinity for the unknown molecules compared to their corresponding initial-specific set (Figure 6C-D). We obtained 23,488 unknown generated molecules with a Glide gscore below -8, of which 11,195 had a maximum similarity against the initial-specific set below 0.3. In contrast, only 1,185 molecules from the initial-specific set met the same threshold. This represents a 20-fold increase from the initial-specific set to the unknown generated molecules. Similarly, we obtained 125 unknown generated molecules below the -10 Glide gscore threshold, compared to only 1 in the initial-specific set. Of these 125 molecules, 64 had a maximum similarity below 0.3 against the initial-specific set.

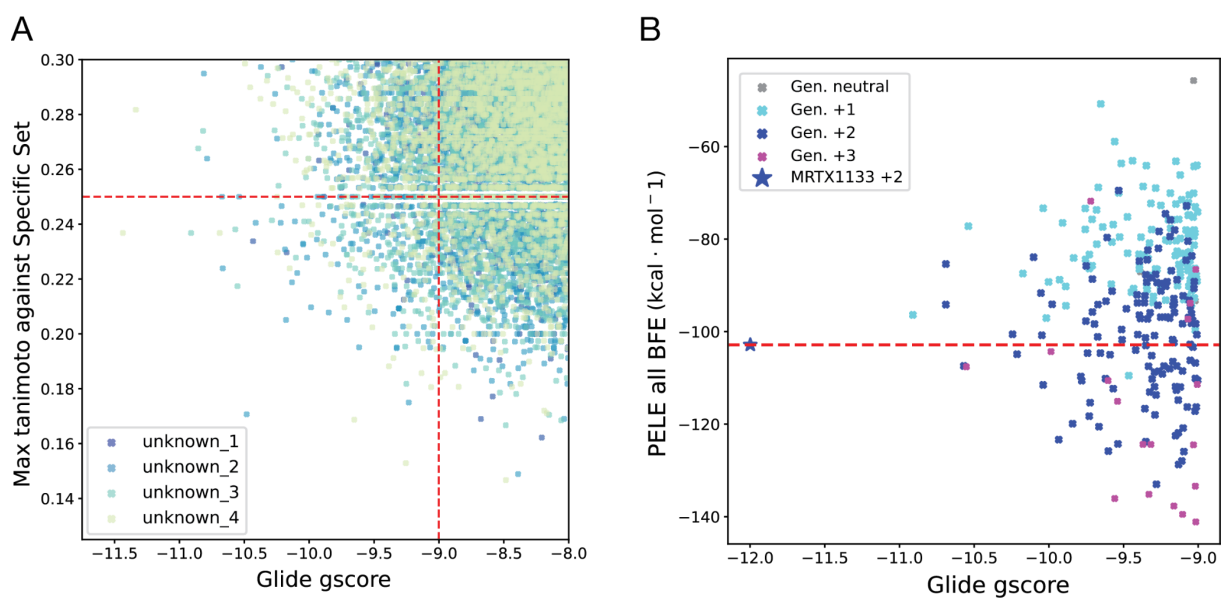

**Figure 7. Selection of KRAS[G12D] candidate inhibitors from the pool of generated molecules with the unknown initial-specific set. A.** Scatter plot of all KRAS[G12D] generated molecules with a Glide gscore below -8 and a maximum similarity against the unknown initial-specific set of 0.3. The red dotted vertical and horizontal lines represent the selection thresholds of -9 for gscore and 0.25 for maximum similarity against the unknown initial-specific set, respectively. **B.** Scatter plot of PELE BFE and Glide gscore of molecules fulfilling panel A's gscore and similarity selection thresholds. Molecules with neutral, +1, +2, and +3 net charges were displayed with a grey, light blue, and dark blue x-filled marker, respectively. Mirati's MRTX1133 molecule protonated at one or two amine groups was displayed with a light blue and a dark blue star marker, respectively.

In line with our approach to CDK2 generated molecules, we conducted a more stringent pose similarity and gscore check with thresholds of 0.25 and -9, respectively (Figure 7A). Additionally, we visually inspected the selected molecules to exclude non-natural chemical structures such as those found in the known generated molecules. This process yielded 255 unknown molecules that met our selection criteria. Subsequently, and based on the enrichment analysis observed in CDK2 (as well as in multiple PELE induced fit refinement studies (31, 32)), we conducted only PELE rescoring simulations on each of these molecules. Based on the formation of two disulfide bonds between ASP12, GLU62, and the two positively charged bicycles of MRTX1133, we scrutinized the net charge of our generated molecules along with their PELE BFE and Glide gscore (Figure 7B). As expected, PELE ranked molecules with a higher net charge more favorably in terms of BFE.

Importantly, we identified a few molecules with a net charge of 3 that we decided to discard due to concerns about their stability, solubility, and permeability. A high charge could compromise their bioavailability and therapeutic potential. Although we did not generate any molecule with a better gscore than the one calculated for MRTX1133 (-11.99), we could obtain 55 molecules with a lower BFE than MRTX1133 (-102.86 kcal·mol$^{-1}$).

# Discussion

We have developed an active learning generative workflow using VAEs and multiple molecular modeling techniques. The main purpose of this is to overcome applicability domain issues from a standalone machine learning procedure: being able to come out of the training molecular space while still producing valuable hits. For this, we centered on two main aspects. First, we emphasized on learning how to make drug-like molecules with good synthetic properties on a relatively different molecular space from the specific training set (by ensuring different enough similarity). Second, we enriched the specific training set with those molecules better performing in molecular docking simulations, aiming at providing better virtual scoring molecules. In practice, these two design steps translated into an iterative procedure involving inner and outer loops. Moreover, the final selection of molecules involves a further hierarchical molecular modelling approach, combining Monte Carlo and MD approaches.

We tested our new GM workflow on two different model systems: CDK2 and KRAS. The choice of CDK2 as the first system was motivated by two main reasons. First, there is extensive experimental data on known CDK2 inhibitors, making it an ideal system for a generative approach. Second, CDK2 could benefit from a generative approach as it could generate novel molecules with unexplored scaffolds. This would help to avoid the typical hinge binder scaffolds of kinases, a potential key aspect to generate specific inhibitors. For this initial system, we ran the entire workflow, obtaining a set of 228 molecules with Glide gscore < -10 and similarity < 0.3. This set was further filtered with PELE induced fit simulations and four final candidates, with original hinge regions, were finally validated with extensive ABFE simulations. Interestingly, the two most expensive modelling techniques, PELE and ABFE, nicely agree on ranking the two "nanomolar predicted" binders (experimental validation of these molecules is on its way). Notice that while PELE introduces great scoring potential, it can run on hundreds of ligands in just a few hours, thus being an ideal filtering scheme to be applied in final selection stages.

The KRAS study aimed to understand how our GM workflow works when presented with only a few known binders in the specific set. Moreover, in this particular case, the known molecules were rather complex, having undergone multiple lead optimization refinement steps and based on just a few cores. Unfortunately, when utilizing these molecules in the specific set, the GM model was unable to produce meaningful molecules. Instead, we opted to use as specific set the results of a virtual screening campaign (thus, using virtual data for the training). This option proved to be more valuable, as it resulted in a diverse set of nicely looking molecules with good predicted scoring values. Interestingly, when starting from each of the sets, we explored quite different molecular spaces; in this sense, the UMAP representation provides a nice visualization of the expanding coverage of the molecular space by our GM workflow, indicating its effectiveness in generating diverse and novel compounds. This last exercise, reinforces the idea of using virtual data for the initial specific set training, a procedure easily implemented on almost any target. Notice that the final list of candidate molecules originates from successive rounds; the initial specific set might be simply an "acceptable" seed to initiate the iterative GM workflow.

While one might argue that using an active learning protocol based on additional modelling tools could further propagate errors, we demonstrate here that the current procedure allows the GM workflow to design molecules with predicted tight binding to protein targets, as reflected in their better docking scores. This is basically what we aimed at: being able to drive the GM workflow to create good looking drug-like molecules with very high docking scores. And not one but multiple scaffolds with good synthesis accessibility scores. Once those values are achieved, we expect to have a significant enrichment factor, on the orders of 30-50%. In this first exercise, we substituted the experimental validation by extensive ABFE molecular dynamics calculations, resulting in two predicted nanomolar binders (out of 4 tested). Similar rates have been achieved at Nostrum Biodiscovery in confidential prospective client studies. Overall, the GM protocol allows to quickly generate dozens of new tight binders, opening the door to different followup strategies involving, for example, novel synthesis or analogue search in billion-like on demand libraries.

# Methods

## Training sets

### General training set

The general training set contains 247,119 SMILES from the ChEMBL 30 compound records database (33). These were selected by applying drug-like filters to the complete database containing more than 2.7 million molecules.

### CDK2 initial-specific training set

The initial-specific training set to CDK2 is composed of 1,061 distinct molecules that have been experimentally tested for their inhibitory potency against CDK2. The inhibitory potency of all these molecules is expressed in $IC_{50}$ values that range from 0.2 nm to 100 μm. These molecules were selected from both ChEMBL (33) and PDBbind (34) . The final selection was diversified for scaffold content.

### KRAS$^{G12D}$ initial-specific training sets

The initial-specific training set of known KRAS$^{G12D}$ inhibitors consists of 60 actual KRAS$^{G12D}$ inhibitors and 13 KRAS$^{G12C}$ inhibitors with trimmed warheads. The inclusion of KRAS$^{G12C}$ inhibitors was based on their binding to the SII allosteric pocket, the same as the KRAS$^{G12D}$ inhibitors. Of the 60 KRAS$^{G12D}$ inhibitors, 10 were acquired from PDB (35). These were obtained by downloading all KRAS$^{G12D}$ crystallographic structures and applying a positional filter to their ligands towards ASP12. The remaining 50 inhibitors were sourced from chemical databases, including 31 from ChEMBL, 14 from PubChem (36), and 5 from MedChemExpress (37). On the other hand, out of the 13 KRAS$^{G12C}$ inhibitors, 7 were obtained from PDB. This time, only KRAS$^{G12C}$ structures were analyzed with a positional filter applied to their ligands towards CYS12. The final 5 inhibitors were either clinically approved or in different clinical phases (38).

The KRAS$^{G12D}$ unknown initial-specific training set is composed of 1,891 molecules. These were obtained from an HTVS on a subset of the Enamine REAL 2022 collection (39). We filtered this subset to include only molecules having 25-32 HAC (acetic acids), 2-3 aromatic rings, and a single basic amine. To perform the HTVS, we used a grid created with the Mirati MRTX1133 molecule and two constraints: 1) a distance constraint between the molecule basic amine and G12D, 2) and a hydrophobic aromatic constraint with the hydrophobic subcavity of SII.

## Target Selection

To select an appropriate target structure for the molecular modelling simulations of the newly generated molecules, we followed these steps:

1) Gathering of candidate target structures: We collected all target X-ray crystallography structures from PDB. The structures were filtered based on specific criteria, such as resolution (<2.5Å) and source organism (homo sapiens).

2) Data processing: We removed all water molecules, ligands, and ions from the candidate target structures and split them into chains. We then discarded chains that fell below a minimum (280 and 155 for CDK2 and KRAS$^{G12D}$, respectively) or exceeded a maximum (300 and 175 for CDK2 and KRAS$^{G12D}$, respectively) number of residues to ensure some level of completeness. Next, all structures were superimposed onto a reference structure using TMalign (40) and prepared with Schrodinger Protein Preparation Wizard (PrepWizard) (41). PrepWizard added all missing hydrogen atoms. Particularly, it used PROPKA (42) to calculate the protonation state of titratable residues at pH 7.0 and ProtAssign to optimize the hydrogen-bond assignments.

3) Binding site volume calculation: We calculated the volume of the ATP-binding site and SII of CDK2 and KRAS$^{G12D}$, respectively, for each candidate target structure using the Schrodinger SiteMap software (43). Next, we generated a pairwise volume overlapping matrix using the *trajectory_binding_site_volumes.py* script from SiteMap.

4) Clustering: We performed hierarchical clustering on the volume overlapping matrix, extracted the resulting dendrogram, and selected clusters based on a threshold using the Seaborn Python library (44). By doing so, we grouped structures based on the similarity of their target binding site. Then, we visually analyzed the differences between clusters and kept only the cluster of interest for the subsequent molecular simulations (e.g., KRAS$^{G12D}$ in its inactive form).

5) Final candidate selection: We defined as the center of the cluster of interest the cluster structure with the highest average overlapping volume with all the structures in the cluster. Next, we extracted the 10 cluster structures with the highest overlapping volume with the cluster center. Eventually, we selected the final candidate target structure from these 10 structures based on several criteria, such as resolution, completeness, quality of the density map, and presence of ligands.

6) ABFE estimation of binding energies: In case of CDK2, the PELE refined poses were subjected to a short MD simulation using the same force field and explicit solvent as the subsequent free energy calculation. In some cases, a docked pose was not stable, and the ligand moved away from the binding site during equilibration. Only poses that remained stable during the short MD simulation (details to be published later) were advanced to ABFE simulation. The ABFE calculations were carried out under the same conditions under which the x-ray ligands for CDK2 was benchmarked. This allowed for reasonably scaling the calculated free energy values as shown in Figure 5C. The KRAS ligands under the current conditions were difficult to accurately estimate using ABFE and thus not reported in the publication

## GM workflow

We utilized several Python libraries to calculate the chemoinformatic metrics required by the inner active learning steps. For the QED, we used the Chem module of the RDKit Python library (45). The QED score falls within a range of 0 to 1, where 0 implies that all druglike properties are unfavorable, and 1 that all properties are favorable (46). We used the DataStructs module of the same Python library to compute molecular similarity with the Tanimoto metric (47) in conjunction with the Morgan4 molecular fingerprint (48). Tanimoto's similarity score ranges from 0 to 1, where 1 indicates that a pair of molecules are identical, and 0 indicates no similarity between molecules. The SA score was obtained with the SAScore module from the Scopy library (49). SA score ranges from 1 to 10, where a lower score indicates that the molecule is easier to synthesize, and a higher score indicates that it is more difficult to synthesize.

# Molecular Modelling

## Glide docking

The Glide protocol used to compute the docking score during the outer active learning step varied depending on the system:

**CDK2:** Given that CDK2's initial-specific set size was sufficient and all molecules in the set had an $IC_{50}$ value, we assessed the predictive capability of various Glide protocols in distinguishing between active and inactive molecules based on their Glide gscore. For this purpose, we used the $IC_{50}$ values as the real classifier and the protocol's Glide gscores as predicted classifiers to compute the area under the curve (AUC) of the ROC curves of each protocol. Specifically, we fixed the $IC_{50}$ threshold at 10µM and varied the gscore threshold within a specified range. For each gscore threshold, we obtained the output of a binary classification, which was then used to plot the ROC curves. By comparing the AUC of the different Glide protocols, we determined which protocol had the highest predictive power in discriminating between active and inactive molecules. The protocol with higher AUC was the XP Glide with a hydrogen bond constraint to the LEU83 of CDK2 (data not shown). Furthermore, the gscore threshold of -8 was determined by selecting all molecules from the initial-specific set of CDK2 with an experimental affinity below 1µm and calculating their average gscore.

**KRAS$^{G12D}$:** In the case of KRAS$^{G12D}$, we could not evaluate the different Glide docking protocols as we did with CDK2 due to the insufficient number of molecules in the known initial-specific set with an associated $IC_{50}$ value. Instead, we assessed different combinations of constraints based on the known binding mode of the MRTX1133 compound (PDB ID 7RPZ) and selected the one that was able to nicely reproduce the original pose found in the co-crystal structure. Additionally, some known inhibitors of KRAS$^{G12D}$, mostly MRTX1133 derivatives (23, 24), were also docked using the same protocol to validate the methodology. Accordingly, we decided to impose two docking constraints to an SP Glide protocol, a hydrogen bond with the ASP12 and a positional hydrophobic constrain centred on the naphthalene binding subcavity

## PELE rescoring

We utilized the PELE software to enhance the granularity of the docking poses and scores of the newly generated molecules that met the specified Glide gscore and similarity thresholds. For each molecule, we conducted a constraint PELE rescoring simulation using the docked structures from Glide as a starting point. A PELE rescoring simulation consists of a local adaptive PELE exploration (50) performed with 32 CPUs running for 20 PELE epochs of 12 Monte Carlo steps. As with the Glide docking simulations, we imposed a hydrogen bond constraint on LEU83 and ASP12 for CDK2 and KRASG12D, respectively. Each simulation took approximately three wall clock hours to complete on the MareNostrum IV supercomputer at the Barcelona Supercomputing Center.

To determine the best possible PELE metric for ranking and selecting final candidate inhibitors, we conducted a test case on the CDK2 initial-specific set. The experiment involved running constraint PELE rescoring simulations on 25 molecules with the highest $IC_{50}$, 25 with the lowest $IC_{50}$, and 50 with intermediate $IC_{50}$ values from the CDK2 initial-specific set. We then extracted 15 PELE metrics based on three different PELE binding energies (BE), which were corrected by the ligand's strain energy, and five different sets of conformations. Then the metrics were correlated with the $IC_{50}$ experimental values, and the one with the highest correlation was selected as the one to be used for CDK2 and KRAS sections. Specifically, we utilized the average BE, minimum BE, and BFE on five sets of conformations, including the complete simulation, the most populated cluster, the cluster with the lowest BE, the cluster with the lowest average BE, and the cluster with the lowest BFE. We found that the BFE obtained from the complete PELE simulation had the highest correlation with $IC_{50}$ (not shown). Therefore, we selected this metric as the PELE metric for ranking and selecting the final candidate inhibitors of CDK2 and KRAS.

Given a set of PELE conformations, its PELE BFE (binding free energy estimator) is defined as the expected value of its BEs,

$$BFE = <BE> = \sum_{i}^{N} p_i BE_i$$

where N is the total number of PELE conformations, $BE_i$ the PELE binding energy at each conformation *i*, and $p_i$ the probability of the $BE_i$ state of energy given by the following Boltzman distribution,

$$p_i = \frac{e^{-TE_i/K_bT}}{Q}$$

where, *TE$_i$* is the total energy of the *BE$_i$* state of energy, *T* the absolute temperature (set at 298K) and *K$_b$* the Boltzman constant. Finally *Q* is defined as:

$$Q = \sum_i^N e^{TE_i/K_bT}$$

Absolute binding free energy simulations:

ABFE simulations were carried out using a double decoupling scheme as described previously in *Boresch et al* (27) and (51). Briefly, starting from the physical ligand in water, the vdw and electrostatic interactions within the ligand and between the ligand and water are slowly turned off until the ligand becomes dummy; second, the dummy ligand is attached to the protein binding pocket through a set of cross-link restraints similar to what was proposed by (*Boresch et al*) in the third step, the intra-ligand and ligand-environment vdw and electrostatic interactions for the restrained ligand are slowly turned on in the protein binding pocket and the cross-link restrains are relaxed after that. Prior to running on ABFE on new ligands, we benchmarked ABFE on an existing set of CDK2 ligands. Further details about the ABFE simulation methods protocol will be released in the full publication.

Data analysis, visualization, and availability

All molecular visualizations (Figure 5C) were rendered with MarvinSketch (52). To generate the UMAP plots (Figure 2D, Figure 3B, and Figure 6B), umap-learn, an open-source Python package (28), was utilized in conjunction with Morgan4 as the chosen molecular fingerprinting (48), and the Hamming Distance (53) as the selected metric. All Python scripts specifically developed for the molecular analysis and visualization developed for this work are available at: https://github.com/IFilella/MolecularAnalysis

# Supplementary

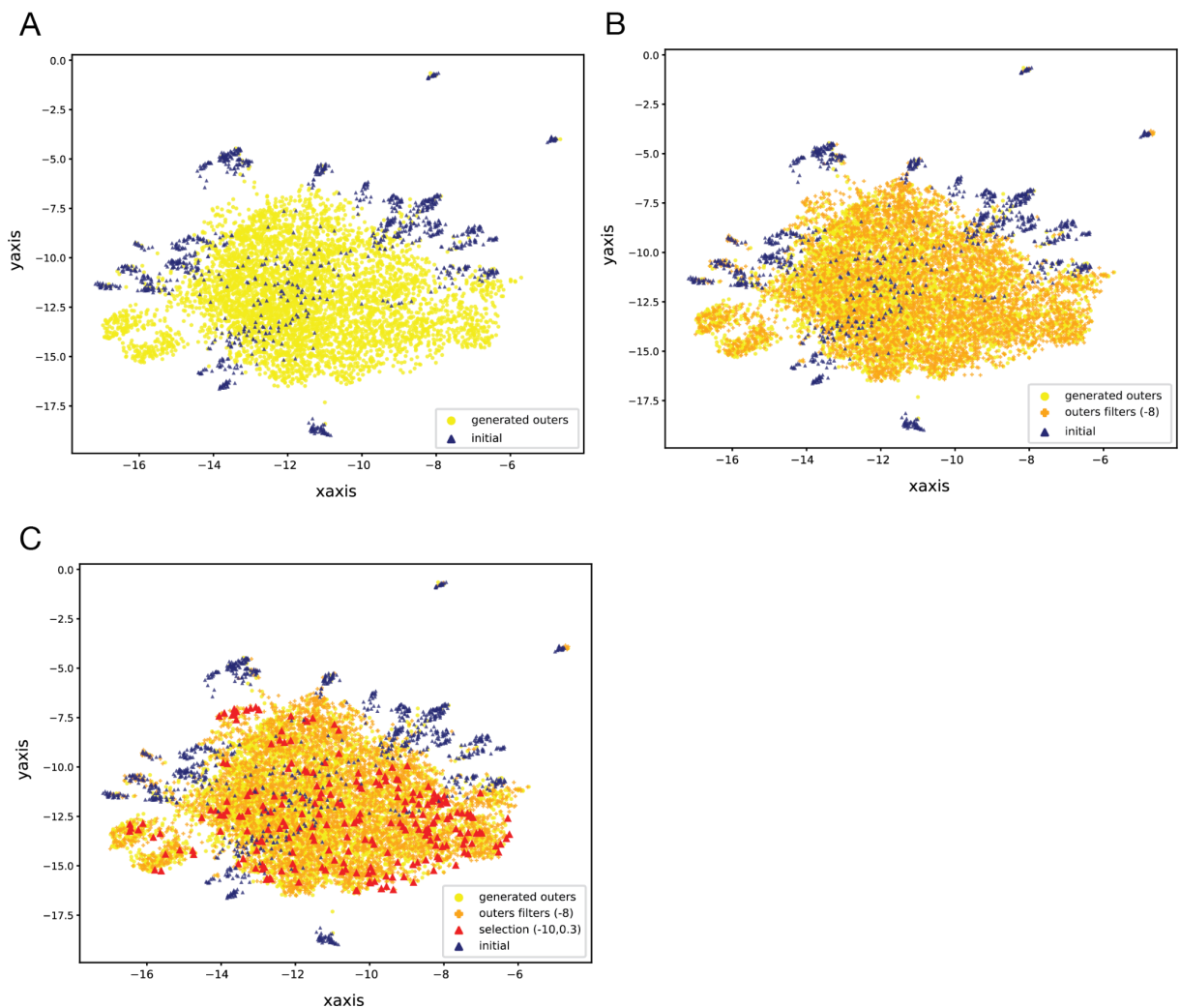

**Supplementary Figure 1. UMAP plot illustrating the GM workflow's complete chemical space exploration of CDK2 inhibitors. A)** UMAP with the generated molecules over the five outer loops and the initial-specific set of CDK2 molecules **B)** UMAP from panel A with the position of the molecules that passed the outer active learning step filters highlighted in orange **C)** UMAP from panel B with the position of molecules meeting the final selection stringent filters highlighted in red.

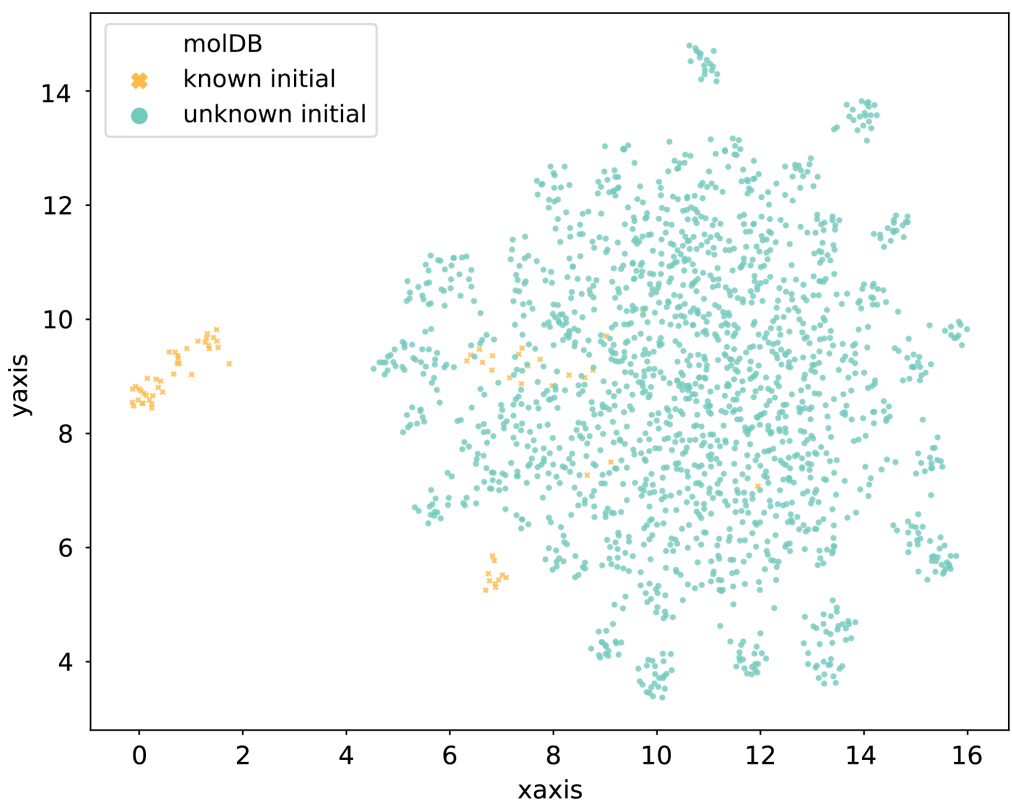

**Supplementary Figure 2.** UMAP plot of the molecules from the KRAS known and unknown specific set.

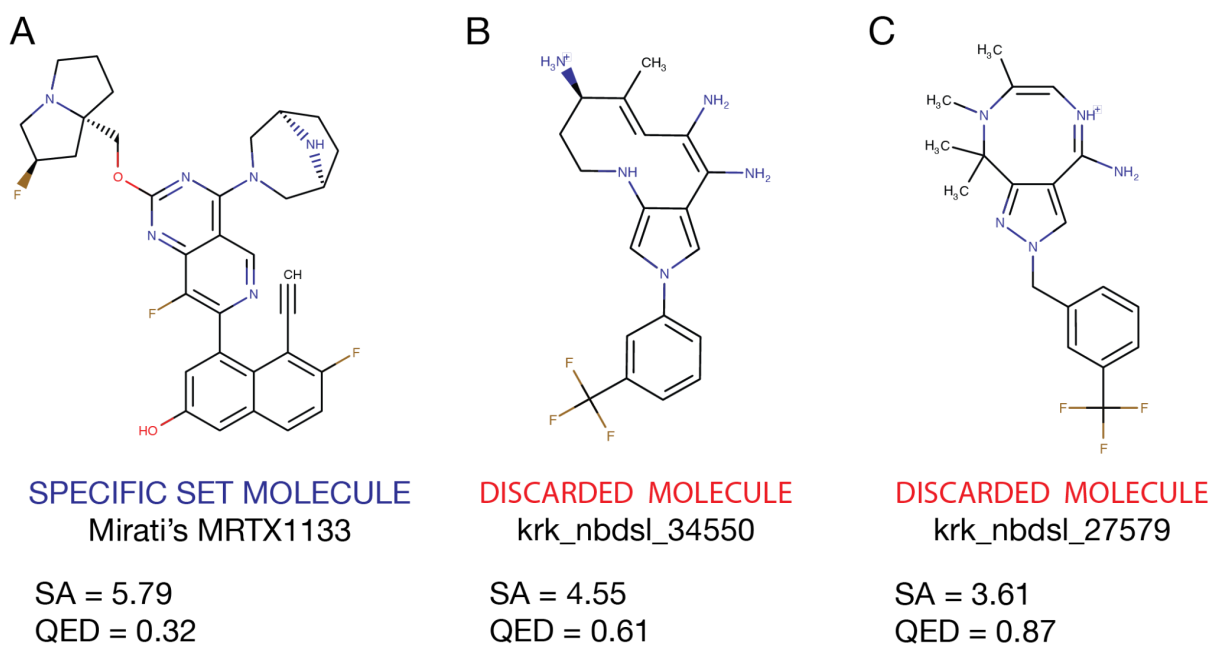

**SPECIFIC SET MOLECULE**
Mirati's MRTX1133

SA = 5.79
QED = 0.32

DISCARDED MOLECULE
krk_nbdsl_34550

SA = 4.55
QED = 0.61

DISCARDED MOLECULE
krk_nbdsl_27579

SA = 3.61
QED = 0.87

**Supplementary Figure 3. MRTX133 Mirati's inhibitor and resulting artifacts from the known generated molecules. A)** MRTX1133 KRAS$^{G12D}$ inhibitor. **B)** krk_nbdsl_34550 known generated molecule exemplifying a discarded molecule due to a broken bicycle **C)**

krk_nbdsl_27579 known generated molecule exemplifying a discarded molecule due to a 8-membered ring. Based on the SA score of MRTX133, which belongs to the initial set of known inhibitors, we set our SA threshold at 7. This threshold was selected to ensure that molecules with a chemistry similar to an FDA approved drug passed the inner active learning step. Unfortunately, this decision increased the number of molecules with dubious synthetic accessibility in the known generation.

|  | Outer1 | Outer2 | Outer3 | Outer4 | Outer5 |
|---|---|---|---|---|---|
| Generated | 49796 | 24766 | 26387 | 19581 | 36465 |
| Inner active learning | 15068 (30.3%) | 12040 (48.6%) | 2065 (7.8%) | 1446 (7.3%) | 18645 (51.1%) |
| Outer active learning | 885 (1.8%) | 1192 (4.8%) | 167 (0.6%) | 88 (0.5%) | 2295 (6.3%) |

**Supplementary Table 1.** The total number of generated molecules and molecules passing inner and outer active learning steps for each outer loop of CDK2 generation.

|  |  | Outer1 | Outer2 | Outer3 | Outer4 |
|---|---|---|---|---|---|
| Known | Generated | 47875 | 42039 | 40303 | 23248 |
| Known | Inner active learning | 9646 (20.1%) | 13370 (31.8%) | 14199 (35.2%) | 13481 (57.9%) |
| Known | Outer active learning | 387 (0.8%) | 982 (2.3%) | 1758 (4.4%) | 408 (1.7%) |
| Unknown | Generated | 37143 | 34452 | 50127 | 30343 |
| Unknown | Inner active learning | 16690 (44.9%) | 15543 (45.1%) | 17806 (35.6%) | 12386 (40.8%) |
| Unknown | Outer active learning | 4432 (11.9%) | 6175 (17.9%) | 7082 (14.1%) | 5799 (19.1%) |

**Supplementary Table 2.** The total number of generated molecules and molecules passing inner and outer active learning steps for each outer loop and each KRAS initial-specific set (Known and Unknown).